\documentclass[journal]{IEEEtran}
\usepackage{amsmath}
\usepackage{multicol}
\usepackage{comment}
\usepackage{multirow}
\usepackage{graphicx}
\usepackage{epstopdf}
\usepackage{color}
\usepackage{diagbox}

\usepackage{ gensymb }
\usepackage[english]{babel}
\usepackage{latexsym}
\usepackage{times}
\usepackage{graphicx}
\usepackage{booktabs}
\usepackage{float}
\usepackage{CJK}
\usepackage{mathtools,amsmath, amsfonts, amssymb, amsthm,color,bm}
\usepackage{cite}
\usepackage[figurename=Fig.,font={small,it}]{caption}
\usepackage{algorithm}
\usepackage{algorithmic} 
\usepackage{siunitx}
\def\N{{\mathcal{N}}}
\def\C{{\mathbb{C}}}
\def\S{{\mathcal{S}}}
\def\R{{\mathbb{R}}}

\title{  Real-time   Faulted Line Localization and {PMU} Placement in Power Systems through Convolutional Neural Networks}

\author{
	\IEEEauthorblockN{Wenting Li, \textit{Student Member, IEEE}, Deepjyoti Deka, \textit{Member, IEEE}, \\
	 Michael Chertkov, \textit{Senior Member, IEEE}, Meng Wang, \textit{Member, IEEE},}
	\thanks{W.~Li, M.~Wang are with the Dept. of Electrical, Computer, and Systems Engineering, Rensselaer Polytechnic Institute,  Troy, NY. Email: \{liw14, wangm7\}@rpi.edu.
	
D.~Deka	 and M.~Chertkov are with the Theory Division and the Center for Nonlinear Studies, Los Alamos National Laboratory, Los Alamos, NM.
Email: \{deepjyoti, chertkov\}@lanl.gov.
This research is supported in part by ARO W911NF-17-1-0407 and the Center for Future Energy Systems (CFES) and the Department of Energy through the Grid Modernization Lab Consortium and the Center for Non-Linear Studies (CNLS) at Los Alamos National Laboratory. }
}
 
\begin{document}
\maketitle
\begin{abstract}
Diverse fault types, fast re-closures, and complicated transient states after a fault event make real-time fault location in power grids challenging. Existing localization techniques in this area rely on simplistic assumptions, such as static loads, or require much higher sampling rates or total measurement availability. This paper proposes a faulted line localization method based on a Convolutional Neural Network (CNN) classifier using bus voltages. Unlike prior data-driven methods, the proposed classifier is based on features with physical interpretations that improve the robustness of the location performance. The accuracy of our CNN based localization tool is demonstrably superior to other machine learning classifiers in the literature. To further improve the location performance, a joint phasor measurement units (PMU) placement strategy is proposed and validated against other methods. A significant aspect of our methodology is that under very low observability ($7\%$ of buses), the algorithm is still able to localize the faulted line to a small neighborhood with high probability. The performance of our scheme is validated through simulations of faults of various types in the IEEE 39-bus and 68-bus power systems under varying uncertain conditions, system observability, and measurement quality. 
\end{abstract} 
\section{Introduction} 
System restoration in large-scale power grids needs methods to locate and isolate the faulted lines efficiently. While circuit breakers and relays are widely applied to check the status of transmission lines, their mis-operation occasionally happens and has in the past caused severe cascading failures or blackouts, as reported in the 2003 blackout analysis in America \cite{NBHmTWS10}. This motivates the development of intelligent monitoring systems in addition to the traditional methods. Among others, data-driven methods for monitoring and localization now can be implemented due to the placement of wide-area measurement devices, like phasor measurement units (PMUs) in the North American grid, that generate high-resolution data \cite{WCG15}. 

Fault location methods are used to pinpoint the fault point in the faulted lines. Prior methods in this context can be categorized into two classes: 1) conventional, and 2) wide-area algorithms. The conventional algorithms for use in one-end, double-end and multi-end transmission lines include impedance-based, traveling-wave based and artificial intelligent methods \cite{FSMH17,JZGWZ14,FSS18,LJ51,HLH17,YSA15}. The impedance-based method computes the fault currents according to the measured voltages and known bus impedance and then determines the fault distance to the measured point using distributed parameters \cite{ FSMH17,JZGWZ14,FSS18 }. Such methods assume static loads and are sensitive to the values of line parameters. Traveling-wave based approaches have been applied to locate faults in the power system since the 1950s, by relating the time difference of traveling-wave propagating to the terminal buses to localizing the fault point. These approaches require high accuracy of time synchronization and a high sampling rate of measurements, and that restrict their usage \cite{LJ51,HLH17}. Finally, artificial intelligence based methods train neural networks (NN) or support vector machines (SVM) with the extracted features from collected measurements. Generally, the raw data need to be pre-processed through Fourier or wavelet transformation, and the coefficients or fundamental components of the transformed data are treated as features \cite{LEY13,YSA15}. Due to the transformation, the requirement of sampling rate is high, like 1kHz, which is far beyond the capability of PMUs. Further, the majority of conventional methods in the literature require at least one terminal buses of each line to be measured, and hence are not suitable for large-scale power system where only a limited number of PMU are installed. To overcome the issues of localizing faults with non-ubiquitous PMUs, wide-area algorithms have been proposed using information collected, not just locally but from the overall geographical network. 
 
For the wide-area methods, the faulted line is first localized (either exactly or approximately), and then the exact position along the line is determined at a second stage \cite{APM15,FAA16, MAFS17,KAAA12}. The first stage itself is challenging due to the large number of candidate lines and limited measurements. In \cite{APM15}, the residual between the estimated and measured voltages variations is computed for each transmission line to determine the one with minimum residual (faulted line), and then the positive and negative network parameters are employed to further determine the position of the fault point. Note that this method is computationally expensive as all lines need to be examined. The authors in \cite{MAFS17,FAA16} largely reduce the computational complexity by formulating the search process into a sparse optimization problem, and the values larger than a pre-defined threshold indicate the faulted area. These methods achieve small fault distance errors with high-resolution measurements (more than 2000 Hz) under the low level of noise, and the robustness to random load fluctuations and the inaccurate voltage measurements is lack of analysis. 

 In addition, traveling-wave based methods using wide-area measurements have been proposed, but their high sampling rate (1 MHz) increases the cost of installing devices in a large-scale system \cite{KAAA12}. It is worth mentioning that a parallel line of work on line outage localization for state estimation or topology estimation \cite{ZG12} based on direct current (DC) power flow models exist. Such works, however, assume small load variations and only employ steady state measurements, and hence have restricted usage compared to our and other works that study real-time localization with transient measurements.  

This paper concentrates on the first stage of fault localization problem, i.e., on the issue of locating the faulted lines or faulted area, in particular in the regime of limited measurement availability. This paper defines the \textit{feature vector} according to the substitution theory, and design a four-layer CNN classifier to locate the faulted line, and then propose a joint PMU placement algorithm to improve the location performance further. The main contributions are summarized from four aspects: 1) The \textit{feature vector} with clear physical interpretations is defined (Section \ref{interpret}). With this feature as input, the complexity of the CNN classifier is largely reduced, compared with the classic ones with millions of parameters \cite{KSHE12,SVLV14}. Moreover, faults can be located within the small neighborhood of the faulted point when a limited number of buses are measured (Section \ref{arc} and \ref{neighbor}). 2) The four-layer CNN classifier is motivated due to the property of sparse connectivity and capturing local features (Section \ref{method}). The advantages of CNN are illustrated by the comparison with other types of classifiers (Section \ref{comp}). 3) A PMU placement algorithm is proposed based on the loss function of the CNN classifier, and its localization performance is shown to be improved compared with other random or topology-based methods (Section \ref{PMUs} and \ref{cmp_pmu}). 4) The significance of the localization scheme lies in its robustness to practical scenarios with differing load variations (Section \ref{load}), noise (Section \ref{noise}), and voltage quality (Section \ref{quality}).

The remaining parts are organized as follows: the two main components of the proposed method are physically interpretable feature extraction and CNN based classification, respectively explained in Section II and III. The structure of the CNN is described, and the motivations are itemized. Section IV explains the proposed algorithm to utilize the CNN classification into determining the PMU placement. The numerical results in Section V validate the proposed method both in the IEEE 39-bus and 68-bus power systems. Then Section VI concludes results and discusses future research.  
% \large{ } 
\begin{table} %[!h] 
\vspace{2.5mm}
\caption*{ \centering{NOMENCLATURE}} 
 \begin{tabular}{l l } 
 $U^0, I^0$ & Bus voltages and currents before faults \\ [0.5ex] 
 $U^{\prime}, I^{\prime}  $ & Bus voltages and currents during faults \\ [0.5ex]
$U_{f}, I_{f} $ & The voltage and current of the fault point \\ [0.5ex] 
 $Y^{0}, Y^{\prime}  $ & Admittance matrix before and during faults \\ [0.5ex]
 $Y^{F} $ & Admittance matrix including the fault point during faults \\ [0.5ex] 
 $\Delta I^u $ & The unbalanced current \\ 
 $\psi^j $ & The $j$th feature vector when all buses are measured \\ [0.5ex]
 $ \bar{\psi}^j$ & The $j$th feature vector when partial buses are measured \\ [0.5ex] 
 $y^j$ & The label of the $j$th dataset \\ [0.5ex]
 $ W_k, C_k $ & The  kernel and output of the $k$th convolutional layer  \\ [0.5ex]
 $ R_k$ & The output of the $k$th Rectified Linear Unit (ReLU) layer \\ [0.5ex]
 $ P_k$ & The output of the $k$th pooling layer \\ [0.5ex]  
  $ \vec{P} $ & The vectorization of the output of the last pooling layer \\ [0.5ex]
 $ \bar{y} $ & The estimated probability of all the possible lines by the CNN \\ [0.5ex]   
 $ W_o, B_o $ & The weight and bias matrices of the output layer \\ [0.5ex] 
% $ g(x) $ & The softmax activation function \\ [0.5ex]  
$ \Theta $ & The set of all the parameters learned by the CNN \\ [0.5ex]  
$ \lambda $ & The regularization coefficient \\ [0.5ex] 
$ \S $ & The set of measured buses\\ [0.5ex] 
$K $ & The number of measured buses\\ [0.5ex] 
$ d_i $ & The degree of the $i$th bus\\ [0.5ex] 
$ \beta $ & The weight coefficient\\ [0.5ex] 
$ W_{NN}^i $ & The weight matrix of the $i$th layer of NN\\ [0.5ex]
$ B_{NN}^i $ & The bias matrix of the $i$th layer of NN\\ [0.5ex] 
\end{tabular} 
\vspace{-5mm}
\end{table}   
\section{Feature Extraction for Faulted Line Localization}\label{model}
\begin{figure}[!ht]
	\centering
	\includegraphics[width=0.48 \textwidth]{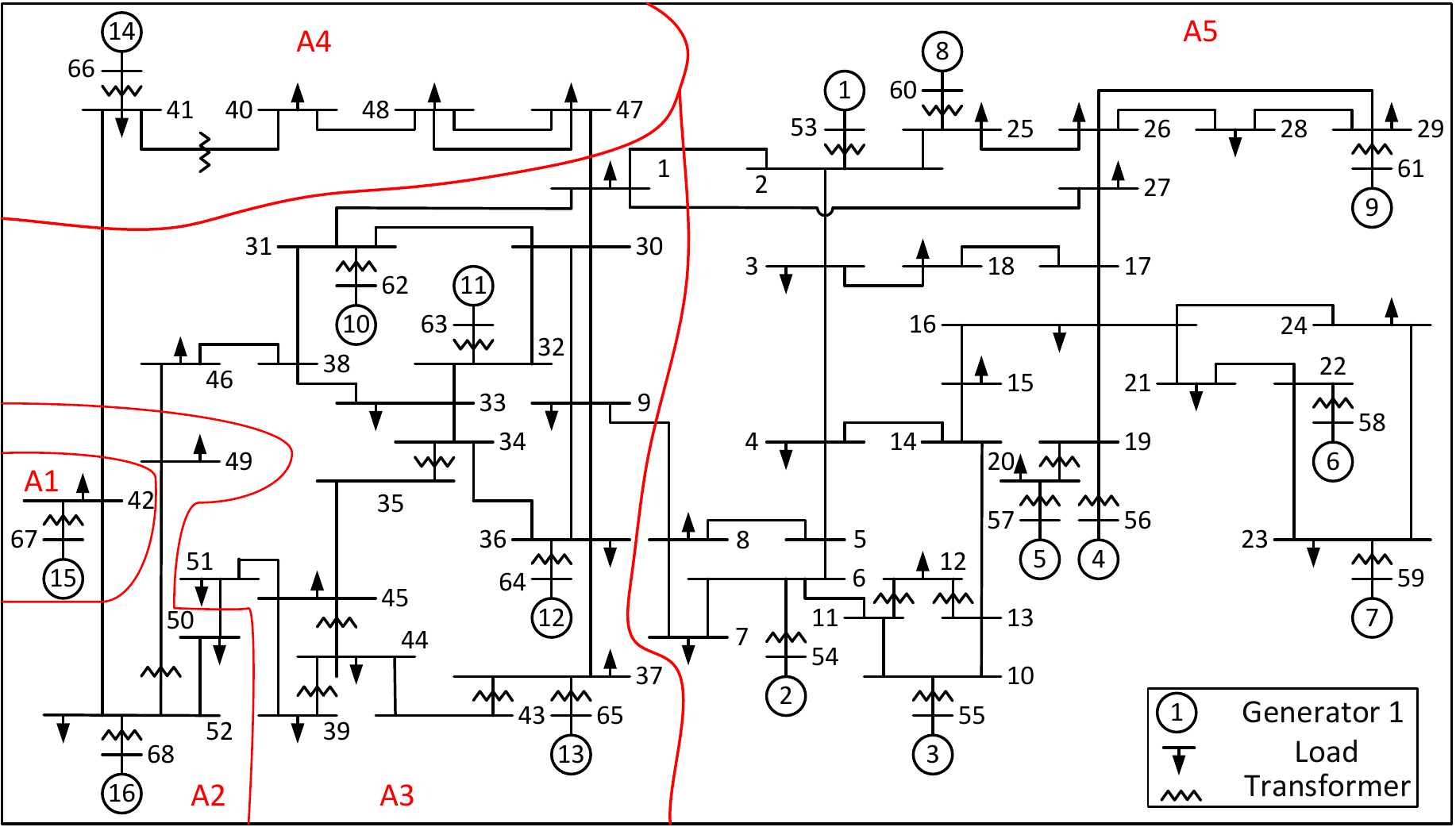} 	
	\caption{IEEE 68-bus system with five coherence groups \cite{RG12}.} \label{network}
		\vspace*{-3mm}
\end{figure}

We consider a power grid of $n$ buses (see Fig. \ref{network}) with a single line fault that may either be one of the following: three-phase short circuit (TP), line to ground (LG), double line to ground (DLG) and line to line (LL) faults. Assuming that fault detection through known techniques \cite{XCLR14} is successful, we are interested in real-time localization of the faulted line using PMU measurements collected before and during the fault from a subset of the grid buses. The main idea is to extract features that characterize the location, and then classify all possible lines with the features. %To this end, we propose to use a convolutional neural network based fault localization method using power-system features derived from the collected data. %As mentioned in the Introduction, selection of right features play a critical role in the success of data-driven classification methods. 
We now describe the selection of the physical model driven feature vector $\psi$, first under complete and then under partial system observability. 
\textbf{Note}: Vectors are marked as bold font or $\vec{\cdot}$ and the real number and complex number sets are respectively represented by $\R$ and $\C$. 
\vspace{-5mm}
\subsection{ Substitution Theory and Features for Full Observability}
In the case of a $n$-bus power system without un-transposed lines\footnote{The un-transposed lines have different mutual impedance between buses and are beyond our analysis.}, we apply the substitution theory \cite{JWLX14} to derive the equations related to pre and during-fault system variables. Given that three-phase measurements may not be available from all the meters, we use only positive sequence data to represent the quantities.

In the steady state regime prior to the fault, bus voltages $ U^0 \in \C^{n \times 1} =[U_1^0, \cdots, U_{n }^0]^T$, currents $I^0 \in \C^{n \times 1} = [I_1^0, \cdots, I_n^0]^T$ and bus admittance matrix $Y^0 \in \C^{n \times n}$ satisfy the Ohm's law in \eqref{pre}, where the $j$th entry in the $i$th row of $Y^0$ is $Y^0_{ij}, i, j = 1, \cdots, n$, denoting the admittance between the bus $i$ and $j$,.
\begin{align}\label{pre}
I^0 & = Y^0 U^0
\end{align}
When the line between the bus $i$ and $j$ is faulted at point F, the during-fault admittance matrix, $Y^F \in \C^{(n+1) \times (n+1)}$, with the fault point F as the $(n+1)$th node can be constructed as
\footnotesize{
\begin{align}
Y^F & =
\begin{bmatrix}
 \begin{array}{@{}c|c@{}}
 \begin{matrix}
 Y_{11} & \cdots & \cdots & \cdots & Y_{1,n } \\
 \cdots & \cdots & \cdots & \cdots & \cdots \\
 \cdots & Y_{ii}^{\prime} & \cdots & Y_{ij}^{\prime} & \cdots \\
 \cdots & \cdots & \cdots & \cdots & \cdots \\
 \cdots & Y_{ji}^{\prime} & \cdots & Y_{jj}^{\prime} & \cdots \\
 \cdots & \cdots & \cdots & \cdots & \cdots \\
 Y_{n,1} & \cdots & \cdots & \cdots & Y_{n,n} \\
 \end{matrix}
 & \textbf{y}_{f_1} \\
 \cmidrule[0.4pt]{1-2}
 \textbf{y}_{f_1}^T & \textbf{y}_{f_2} \\
\end{array}
\end{bmatrix} = \begin{bmatrix}
 \begin{array}{c|c}
 Y^{\prime} & \textbf{y}_{f_1} \\
 \hline
 \textbf{y}_{f_1}^T & \textbf{y}_{f_2}
 \end{array}
 \end{bmatrix},
\end{align}}
\normalsize
where $Y^{\prime}\in \C^{n \times n }$ is the during-fault admittance matrix of $n$ buses, $\textbf{y}_{f_1} = [Y^F_{1,n+1}, \cdots, Y^F_{n,n+1}]^T \in \C^{n \times 1 }$ is the admittance between the F and other buses, $\textbf{y}_{f_2} = Y^F_{n+1,n+1} \in \C $ is the self-admittance of the faulted point F.

During-fault current and voltage $I^{\prime} \in \C^{n \times 1}, U^{\prime} \in \C^{n \times 1} $ of buses, the fault point current and voltage $I_f,U_f$ satisfy 
\begin{align}\label{dur}
\begin{bmatrix}
I^{\prime}\\
I_f
\end{bmatrix} & = Y^{F} \begin{bmatrix}
U^{\prime} \\
U_f
\end{bmatrix} = \begin{bmatrix}
 \begin{array}{c|c}
 Y^{\prime} & \textbf{y}_{f_1} \\
 \hline
 \textbf{y}_{f_1}^T & \textbf{y}_{f_2}
 \end{array}
 \end{bmatrix} \begin{bmatrix}
U^{\prime} \\
U_f 
\end{bmatrix}, \\
\Rightarrow I^{\prime} & = Y^{\prime} U^{\prime} + \textbf{y}_{f_1} U_f 
\end{align}
Replacing the $Y^{\prime}$ by $Y^{\prime} =Y^0 - Y^u$, where $Y^u$ is a 4-sparse\footnote{$k$-sparsity means there are only $k$ nonzero entries.} matrix that only has four nonzero entries $Y^u_{ii} = Y_{ii} - Y^{\prime}_{ii}, Y^u_{ij} = Y_{ij} - Y^{\prime}_{ij}, Y^u_{ji} = Y_{ji} - Y^{\prime}_{ji}, Y^u_{jj} = Y_{jj} - Y^{\prime}_{jj} $, we obtain
\begin{align}
I^{\prime} & = (Y^0 - Y^u) U^{\prime} + \textbf{y}_{f_1} U_f = Y^0 U^{\prime} - \Delta I^u \label{dur_final}
\end{align}
where the \textit{unbalanced current} $\Delta I^u = Y^u U^{\prime} - \textbf{y}_{f_1} U_f $ is a 2-sparse vector with nonzero entries $ \Delta I^u_i, \Delta I^u_j$ given in \eqref{un}. Notice that these nonzero entries are just the terminal buses $i,j$ of the faulted line.
%= [0, \cdots, \Delta I^u_i, \cdots, 0, \Delta I^u_j, \cdots, 0]
 \begin{align} \label{un}
 \Delta I^u_i & =
 (Y_{ii}^{\prime} -Y_{ii}^0 ) U_i^{\prime} + ( Y_{ij}^{\prime}-Y_{ij}^0)U_j^{\prime} - Y_{i(n+1)}^{\prime}U_{n+1}^{\prime} \\
 \Delta I^u_j &= ( Y_{ji}^{\prime}-Y_{ji}^0) U_i^{\prime} + ( Y_{jj}^{\prime}-Y_{jj}^0 )U_j^{\prime} - Y_{j(n+1)}^{\prime}U_{n+1}^{\prime}
 \end{align}
If we define variations of voltage and current as $\Delta U = U^{\prime} - U^0 $, $\Delta I = I^{\prime} - I^0 $ and combine \eqref{pre} and \eqref{dur_final}, then their relationships with the pre-fault admittance $Y^0$ become:
\begin{align}\label{inter}
Y^0 \Delta U & = \Delta I^u+ \Delta I
\end{align}

The \textit{feature vector} $\psi \in \C^{n \times 1} $ is defined according to \eqref{feature} in terms of the bus voltages variations $\Delta U$ before and during the faults and the admittance matrix $Y^0$ before the faults
\begin{align}\label{feature}
\psi & = Y^0 \Delta U.
\end{align}

Because both imaginary and real parts of $ \psi $ can reflect the location, and the imaginary parts show a better performance in a large number of classification experiments, we choose the imaginary part
%$\psi = [\text{Real}(\psi_1), \cdots,\text{Real}(\psi_n)
%[\text{Imag}(\psi_1), \cdots,\text{Imag}(\psi_n) ]$
$\psi$ as the feature input to the classifier to avoid unnecessary complication. %$\psi$ was also used to locate by solving an optimization problem, but usually a threshold needs to be set by experience to determine the final locations \cite{MAFS17, FAA16}.

\subsection{Physical Interpretation of the Features}\label{interpret}
The physical interpretation of $\psi$ is revealed by the two components in \eqref{inter}. The dominant element is $\Delta I^u$, which is a 2-sparse vector with nonzero values exactly corresponding to the terminal buses of the faulted line. Distribution of the $\psi$'s entries is indicative of the faulted line location.

\subsubsection{Feature Extraction under Full Measurements} \label{neigh}
Consider the line between bus $i$ and $j$ as faulted. The $k$th ( $k\neq i, j$) entry $\psi_k$ is not related directly to the faulted line,
\begin{align}
\psi_k & = \Delta I_k + \Delta I_k^u = \Sigma_{l \in \N_k} Y^0_{kl} \Delta U_l = \Sigma_{l \in \N_k} \Delta I_{kl} \label{neighb}
\end{align}
where $\N_k$ denotes the neighbor of the bus $k$, and $ I_{kl}$ is the line currents between the bus $k$ and $l$. Therefore, $\psi_k$ is nonzero if line currents variations in its neighborhood are nonzero. The minor components in $\Delta I$ are therefore useful indicators in the neighborhood of the faulted line. (This conjecture will be post-factum validated below.) 

\subsubsection{Feature Extraction under Partial Observability}\label{sparse}
Assume that only $s < n$ buses in the set $\S$ are measured, and their pre-fault and during-fault voltages are provided, then we derive at the observed buses, $\Delta \bar{U} = \bar{U}^0 - \bar{U}^{\prime}$.  The feature vector $\bar{\psi} \in \C^{n}$ obtained from the data measured by the $s $ buses is defined as:
 \begin{align} \label{bar}
 \bar{\psi} & = \bar{Y}^0 \Delta \bar{U} 
 \end{align}
where $\bar{Y}^0 \in \C^{n \times s} $ denotes the submatrix of the pre-fault admittance matrix between all the $n$ buses, and the $s$ measured buses, $ \bar{U} $ denotes the voltage variations measured by the $s$ buses, and the $k$th entry of $\bar{\psi}$ can be interpreted as 
 \begin{align} 
 \bar{\psi}_k & =  \Sigma_{p \in (\S \cap \mathcal{N}_k}) Y_{kp}^0 \Delta U_p = \Sigma_{p \in \S \cap \mathcal{N}_k} \Delta I_{kp}
 \end{align}
where $ Y_{kp}^0$ denotes the $k$th row and $p$th column of $Y^0$, $\Delta U_p$ is the $p$th entry of $\Delta U$, and $\Delta I_{kp}$ is the line current variations between the buses $k$ and $p$. Thus $\psi$ is a sparse vector, and the nonzero entries of $\psi$ are the buses that are in the neighborhood of the measured buses, where the neighborhood represents those buses directly connected with the measured buses.  
\subsubsection{Numerical Example} 
\begin{figure}[!ht]
	\centering
	\includegraphics[width=0.24 \textwidth]{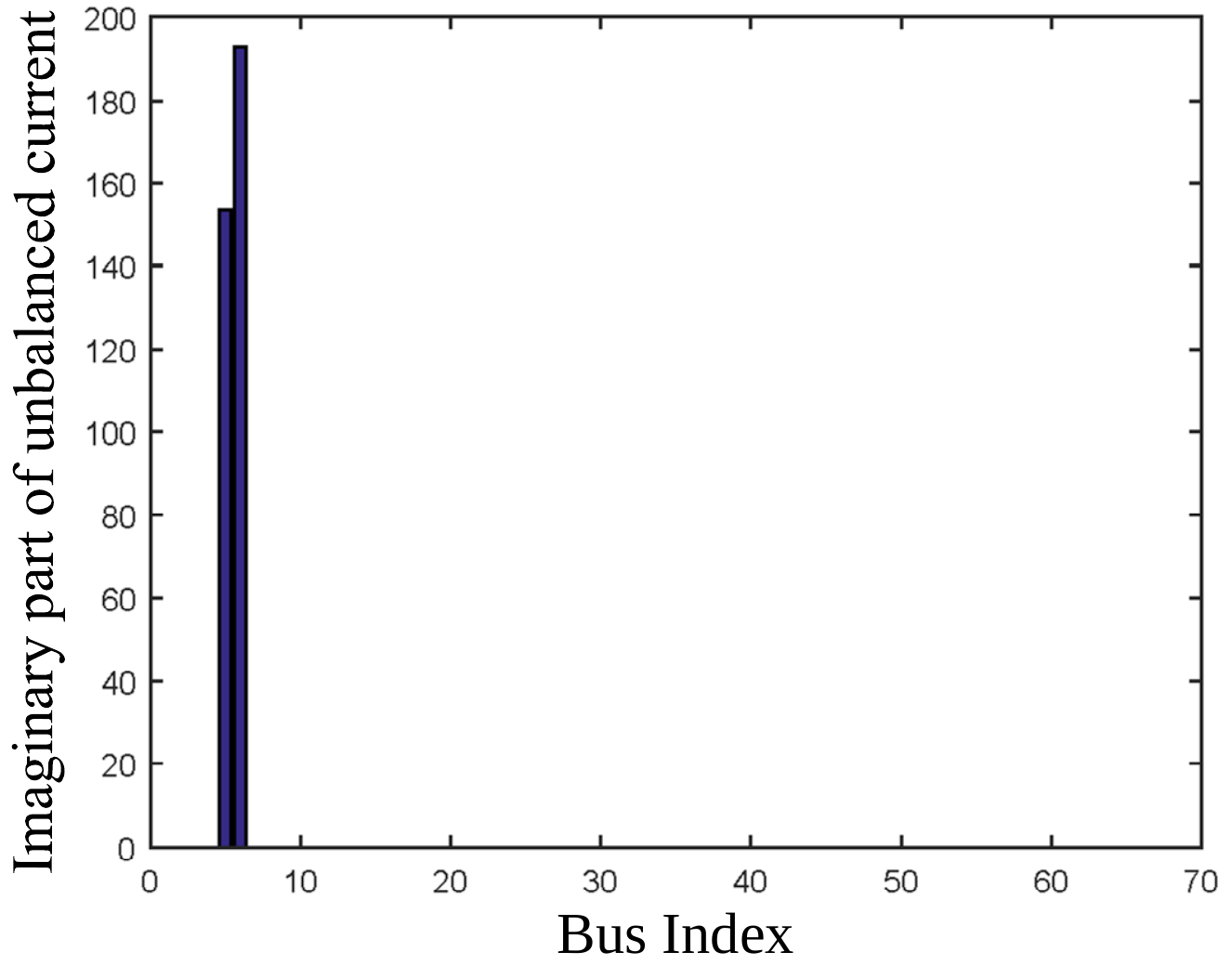} %imag_cur_new.eps
		\includegraphics[width=0.24 \textwidth]{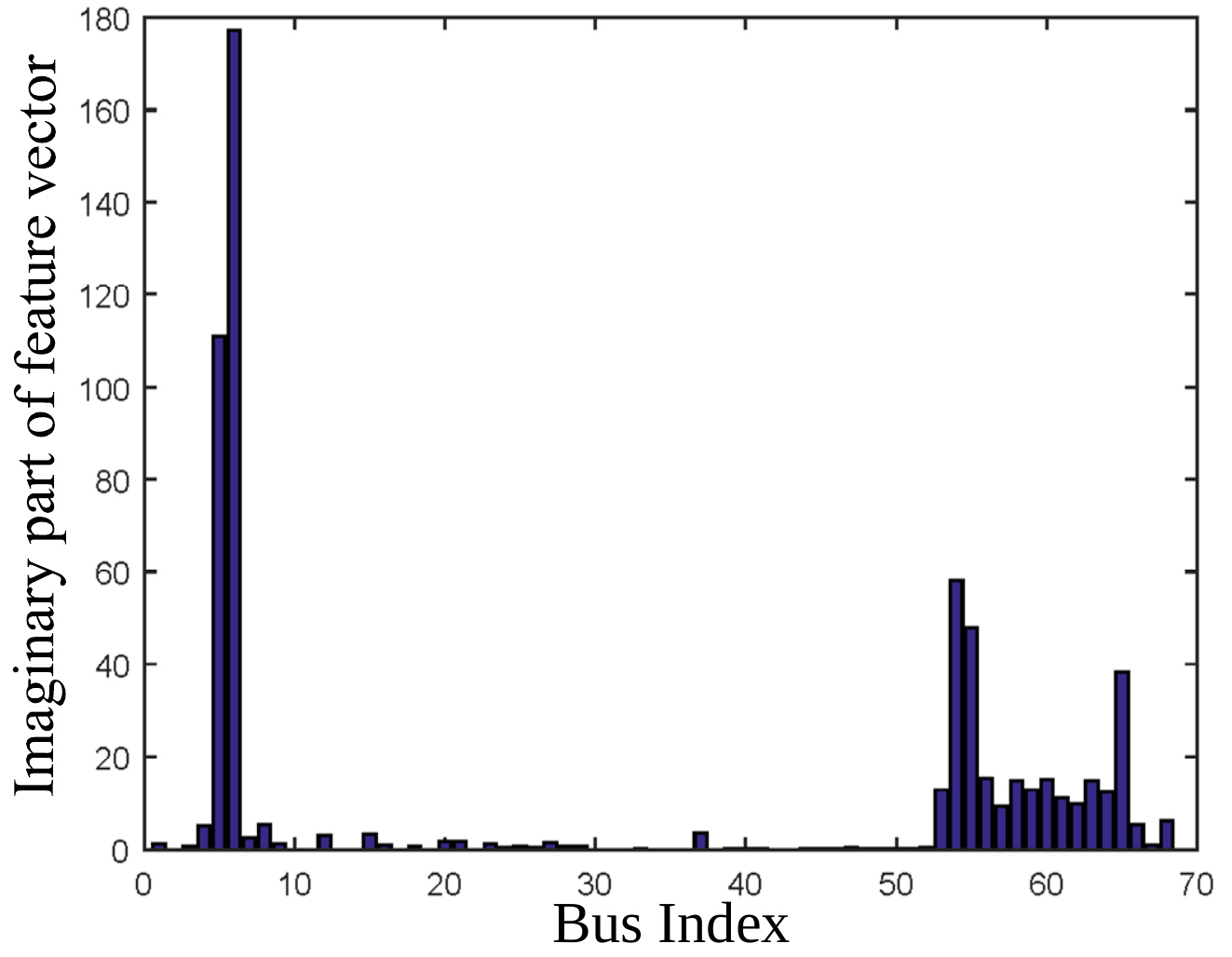} %vari_cur_imag.eps
	\caption{The imaginary parts of the unbalanced currents $\Delta I^u$(left) and of the feature vector $\psi_i, i=1,\cdots, 68$ (right) after a three-phase short circuit fault on the line 5-6 in Fig. \ref{network}} \label{un_cur}
	\vspace*{-3mm}
\end{figure}

We simulate in the power system toolbox (PST), based on nonlinear models \cite{CCW92}, a three-phase short circuit fault lasting 0.2 seconds at the line 5-6 in the IEEE 68-bus power system. The feature vector $\psi$ is computed according to \eqref{feature}. The imaginary parts of $\Delta I^u$ and $\psi \in \C^{68 \times 1}$ shown in Fig.~\ref{un_cur} demonstrate that $\Delta I^u$ is a sparse vector with nonzero entries corresponding to the two terminal buses (5 and 6) of the faulted line, while $\psi_5$ and $\psi_6$ have relatively large values than others. Meanwhile, some other buses have nonzero values as their voltages change during the fault period.%Further, many other buses $(7,8,37,53)$ and $(54-68)$ have nonzero values. These buses are either some PV buses \cite{KPNM94} with large current variations or in the neighborhood of the faulted line.%The former group are close to the faulted line and have changes in line currents, and the latter group of buses are PV buses and hence have large current variations.

The main reason to select $\bar{\psi}$ rather than $\Delta \bar{I}^u$ as the feature vector is that otherwise, measurements of all buses need to be known to ensure the nonzero entries of $\Delta \bar{I}^u$ are included, but in reality not all the buses are measured by PMUs.

 After representing all faults in the dataset by their feature vectors, we label them by their locations. For the system of $m$ lines, we label the dataset into $(m+1)$ classes with the $(m+1)$th class denoting the normal condition. In the next Section, we examine the performance of the classifier.
\vspace{-3mm}
\section{Classification}\label{method}
 Deep/machine learning has produced encouraging improvements in the fields of computer vision \cite{KSHE12}, natural language \cite{SVLV14} and speech recognition \cite{MDHG12}, through the selection of correct data-features. With features $ {\psi}$ extracted for locating faulted lines, many classifiers, e.g., SVM and fully-connected NN, can be applied \cite{RC14}. CNN classifier is preferred considering its property of sparse connectivity and parameter sharing \cite{GBC16}, and the experimental results in Section V-C indicate the advantages of CNN. 
\vspace{-3mm}
\subsection{Architecture Design}\label{CNN_conf}
Although there is no uniform way of designing the structure of CNN, and novel architectures are frequently proposed, several basic components are typically considered together for better classification accuracy in a wide range of applications. These components include convolutional, Rectified Linear Unit (ReLU), Pooling, and fully connected operators. The size of the kernel matrices in these operators and the number of layers are hyper-parameters that are designed to fit the input. In this manuscript, we follow the common practical suggestion - to adopt a scheme which has already shown a competitive advantage in other applications. We build our classifier based on the AlexNet model \cite{KSHE12}. 

We input the imaginary parts of the extracted feature vectors $\psi^j$ and labels $y^j, j=1, \cdots, N$, then the CNN optimizes all the parameters layer by layer.

Let the input of the $k$th convolutional layer ($k =1, \cdots, l$) be $X_k \in R^{w_k \times h_k \times d_k}$, then the feature vector $\psi^j$ of the $j$th dataset is the input of the first layer $ X_1=\psi^j $.
\begin{align}\label{conv}
C^j_k & = X_k \otimes W_k,
\end{align}
where the output of the $k$th convolutional layer is $C^j_k$, which is locally connected with the entries of $X_k$ through kernels $W_k \in R^{c_k, r_k, m_k} $ by the convolution operator $\otimes$ in \eqref{conv} \cite{GBC16}. These kernels element-wise multiply local parts of $X_k$ and also move with the user-defined stride size over the entire input $X_k$. To maintain uniform operations in boundary elements, zeros may be padded to $X_k$.
\begin{align}\label{relu}
R^j_k &= \max(C^j_k, 0)
\end{align}
The convolutional layer is followed by the non-linear ReLU activation function in \eqref{relu}, which discards the negative items of $C^j_k$ without changing the size.
\begin{align}\label{pooling}
P^j_k & = \textit{Pooling}(R^j_k)
\end{align}
To reduce the size of the input at the next layer, the max pooling operator is applied to $R^j_k$ in \eqref{pooling}. Kernels in the pooling operator pick the maximum within a small neighborhood of $R^j_k$ and then move to the next neighborhood with a user-defined stride similar to the convolution operator. Likewise, the user also can pad the $R^j_k$ with zeros to make sizes of the neighborhood and of the kernel equal.

The $P^j_k $ is delivered to the next layer as input $X_{k+1}=P^j_k $. Applying these operators from \eqref{conv} to \eqref{pooling} in all the $l$ layers, the final output $P^j_l$ is vectorized into a long vector $\vec{P}^j $,
\begin{align}\label{out}
\bar{y}^j & = g(W_o^T \vec{P}^j + B_o)
\end{align}
where $W_o, B_o$ are the weight and the bias matrices in the output layer respectively, and $g(\cdot) $ is the softmax function $g(x) = \frac{e^x}{1 + e^x}$. $\vec{P}^j $ is fully connected with the output probability $\bar{y}^j_i, i=1, \cdots, m$ of $m$ lines by \eqref{out}. The line with the highest probability determines the output class or the fault location.
%\vspace*{-1mm}

\begin{figure}[!ht]
	\centering
	\includegraphics[width=0.40\textwidth]{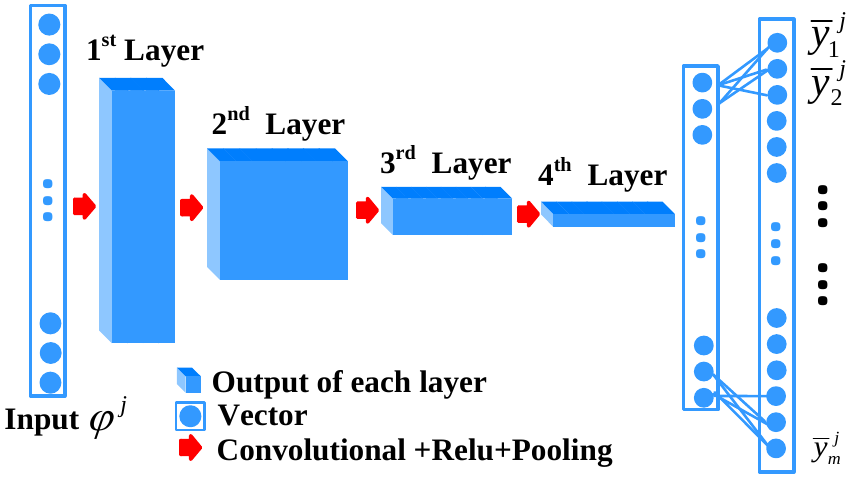}
	 \caption{ The structure of the proposed CNN} \label{CNN}
\vspace{-3mm}	 
\end{figure} 
\vspace{-3mm}	
\subsection{Training Process}
We denote the set of all the CNN parameters $\Theta$. The optimal $\Theta$ is found by minimizing a loss function. Interpreting the output of different classes related to different lines as probabilities of a fault, the cross-entropy loss function \cite{GBC16} together with a regularization term $\lambda \lVert \Theta \rVert_F^2$ to avoid overfitting is the common recipe \eqref{loss}:
\begin{align}\label{loss}
l(\Theta, \S ) & = \frac{1}{N} \Sigma_{j=1}^N \Sigma_{i=1}^n {y}^{j }_i \log{f_{\Theta, \S}( \bar{\psi}^j)} + \lambda \lVert \Theta \rVert_F^2
\end{align}
where $\S$ is the set of measured buses, $\bar{\psi}^j$ is defined in \eqref{bar} with $s \in \S$,
$ y_i^{j } \in R^{m}$ is unity if the label of the $j$-th dataset is $i$, and it is zero otherwise, and $\bar{y}^j_i = f_{\Theta, \S}( \bar{\psi}^j)$ is the output probability of CNN for the fault location of the $j$-th dataset to be at line $i$. $f_{\Theta, \S}(\cdot)$ denotes functions of \eqref{conv} $\sim$ \eqref{out} parameterized by $\Theta$ given the set $\S$ to estimate the probability. $\lambda$ is the regularization coefficient. Notice that the set $\S$ is fixed after the measured buses are determined through some PMU placement methods.  

To solve this optimization problem, the stochastic gradient descent method or some of its extensions like Adam \cite{KBJ14} and RMSprop \cite{HSSK12}, are shown to achieve high classification accuracy in many tests. Although rigorous convergence proofs of gradient-descent based methods are lacking,  many techniques are useful in reducing the effects of initial conditions and also improving the classification accuracy. For example, ``early stopping'' is a convergence criterion to avoid over-fitting, and ``batch normalization'' is effective to the issue of covariance shift \cite{BY12}. 
\vspace*{-2mm}
 \subsection{Motivations of Applying CNN Classifier}
 CNN is preferred mainly considering three reasons: (1) The convolution operation in CNN generates sparse connectivity between the entries of different layers \cite{GBC16}, which is more suitable than the fully connected operation for the sparse vector $\bar{\psi}$ (explained in Section \ref{interpret}).  (2) CNN reveals the local features \cite{GBC16}. The small size of the kernels strengthens CNN to learn the local features instead of the global features. As the neighborhood property of $\bar{\psi}$ (presented in Section \ref{interpret}) illustrates the local characteristics, CNN has the advantages of learning the neighborhood property from the input. (3) In addition, the experimental comparison results (discussed in Section \ref{comp}) of different classifiers further confirm our analysis, and the CNN shows the advantages especially when the system is partially measured.  

In the next Section, we describe how correct PMU placement helps to reduce fault localization error with partial observability and improves the performance of the CNN.
%\vspace{-1mm}	
\section{PMU Placement for Faulted Line Localization under Partial Observability} \label{PMUs} 
\vspace{-1mm}
\begin{algorithm}[!ht]  
\caption{Greedy Algorithm for PMU Placement} \label{alg}
\begin{algorithmic}[1]
\STATE Input: $K, y^j, \bar{\psi}^j, d_i, \beta, \S_{0}, j \in [1,N], i \in [1,n]$
\STATE Initialize : $\S =\S_0, l=\infty$ \\
\WHILE {$|\S|$ is less than $K$ }
\FOR {bus $i \notin \S $}
\STATE Let $\S_i = \{\S \cup i\}$, and compute the loss function $ l_i= \min_{\Theta} l(\Theta, \S_i)$
\ENDFOR
\STATE $ i^* =\arg \min_{i } (\frac{\beta}{d_i} + l_i)$, where $d_i$ is the degree of bus $i$, $\beta $ is a weight coefficient. \label{step7}
\IF {$l_{i^*} < l $}
\STATE $\S = \S_{i^*}, l = l_{i^*} $  
\ENDIF
\ENDWHILE
\STATE Output: $\S $
\end{algorithmic} 
\end{algorithm}   
\vspace{-2mm}	
If the number of PMUs is limited, their correct placement can play a significant role in keeping the quality of the fault localization algorithm described in preceding Section \ref{method}. In this Section, we propose a greedy algorithm to place $K$ PMUs. PMU placement algorithms discussed in the literature, e.g., \cite{YAH12,  MAFS17}, are devised to guarantee full system observability.
However, locating faults may work well with some but not necessarily complete observability. Since the accuracy of localizing the faulted line in our case is determined by the loss function in \eqref{loss}, we suggest optimizing PMU placement (set $\S$) to reduce the loss function of \eqref{loss1}. 
\begin{align}\label{loss1}
\min_{\Theta, \S} & \quad l(\Theta, \S) \\
\text{ s.t. } & |\S| = K \label{st}
 \end{align}
We propose a data-driven placement algorithm that is aware of both the fault localization and the learning mechanism (optimization of the loss function of CNN). To optimize the PMU placement for fault location, the optimal set $\S$ can be obtained by minimizing loss function \eqref{loss1} satisfying \eqref{st} during the training stage.  The regularization item $ \lVert \Theta \rVert_F^2$ reduces the model complexity by creating a parameter set $\Theta$ of a small norm. 
 
To find the optimal set $\S$ of size $K$ is an NP-complete problem, thus we propose an algorithm to greedily increase the number of measured buses until the total number $K$ is reached in Algorithm \ref{alg}. 
 
Given the total number of measured buses $K$, this algorithm greedily increases the size of the set $\S$ from the initial set $\S_{0}$ one by one until $K$, where $\S_{0}$ includes a few buses having the largest degree $d_i$ or being significantly crucial. For each step, the set $\S$ is updated by adding the $i^*$th bus that minimizes the loss function $l_i$ plus the item of $\beta/d_i$. Note that the item $\beta/d_i$ is added to the loss function to account for the effect of grid topology in determining the selected bus. The weight coefficient $\beta \in (0,1)$ adjusts the significance of the bus degree and of the loss function to prioritize the buses with a large degree. This item takes effect obviously when the set $\S$ is large, and the difference of the loss function $l_i$ becomes small.
Meanwhile, a number of experimental results show that adding a bus with more considerable degree tends to have better performances.
Based on all of the above, our algorithm tries to enforce the selected buses to achieve a larger degree by minimizing the loss function augmented with the $\beta/d_i$ item. 
\vspace{-3mm}
\section{Numerical Results} \label{simu}
Four types of line faults, including three-phase short circuit (TP), line to ground(LG), double line to ground (DLG) and line to line (LL) faults are simulated through PST \cite{CCW92}. The initial active and reactive loads $z \in \R^{2n}$ are drawn from the Gaussian distribution $\mathcal{N}(\mu, \varepsilon I)$ with mean $ \mu \in R^{2n}$ and covariance matrix $\Lambda \in R^{2n \times 2n}$, where the mean value of the load $\mu$ is from the standard dataset and the covariance matrix is defined as $\varepsilon I$ with $\varepsilon = 1$. 
The fault impedance changes in the range of 0.0001 to 0.1 per unit (p.u.), and the fault is cleared after 0.2 seconds \cite{KPNM94}. The fault location performance is evaluated by the location accuracy rate (\textbf{LAR}) $\eta$ defined in \eqref{eta}. The data rate of PMU is 60 samples per second.
\begin{equation}\label{eta}
\eta = \frac{\text{The number of faults correctly located}}{\text{total number of faults}}
\end{equation}
 
The proposed approach is validated both in the IEEE 39-bus and the IEEE 68-bus power systems. Two CNN classifiers are constructed for these systems respectively. The LARs of both systems are demonstrated when the test system is partially measured by 15\% to 30\% of buses. The misclassified faults of these two systems are analyzed respectively. Notice that the measured buses are selected by the proposed PMU placement algorithm in Section \ref{PMUs}. Subsequently, more extensive properties of the proposed methods are tested in the IEEE 68-bus power system from Section \ref{comp} to Section \ref{quality}, involving the comparison of different classifiers, comparison of different PMU placement algorithms, and the robustness to uncertainty, low quality of measurements and noise. 
\vspace{-3mm}
\subsection{Performances of 39-bus Test Cases under Partial Observability }\label{39-test}
\subsubsection{Parameters of CNN and Datasets Introduction}
\begin{table}[!h]
\centering
\caption{ The structural parameters of the CNN for the 39-bus power system} \label{CNN39}
\begin{tabular}{c c c  c}
\hline \hline
Layer Type & Operator & Kernel &  Output \\
\hline The $1^{st}$ & Convolution & 4 @ 3 &  4 @ 37\\
 Convolutional & Max Pooling & 2$ \times $1 &  4 @ 19 \\
\hline The $2^{nd}$ & Convolution & 8 @ 3 &   8 @ 17\\
 Convolutional  & Max Pooling & 2$ \times $1 &  8 @ 9 \\
\hline The $3^{rd}$ & Convolution & 8 @ 2 &  8 @ 8\\
 Convolutional & Max Pooling & 2$ \times $1 &  8 @ 4 \\
\hline The $4^{th}$ & Convolution & 8 @ 2 &  8 @ 3\\
 Convolutional & Max Pooling & 2$ \times $1 &  8 @ 2 \\
\hline Fully Connected & Vectorize & - & 16 \\
\hline Output & Regression & - &  47 \\
\hline
\end{tabular}
\end{table} 

The parameters of the CNN classifier built for the IEEE 39-bus power systems are summarized in Table \ref{CNN39}. This CNN has four convolutional layers, one fully connected layer, and one output layer. In Table \ref{CNN39},  ``4 @ 3 '' denotes that there are four kernels of the size 3 by 1;  ``4 @ 37'' denotes that the output is four vectors of the size 37 by 1; ``$2 \times 1$'' denotes that there is one kernel with the dimension of 2 by 1.  In each convolutional layer, in the ``Convolution'' operation the stride size is 1 and zeros are padded; in the ``Max Pooling'' operation, the stride size is 2 and no need to pad zeros. ``-'' denotes that there are no kernels. In the output layer of the CNN in Table~\ref{CNN39}, there are a weight matrix $W_{39}^o \in R^{16 \times 47}$ and a bias matrix $B^o_{39} \in R^{47}$. The number of classes $m=47$, including 46 locations of the faulted line and one null outage line \cite{GCVBLE16}. 

To train the CNN classifiers we set $\lambda = 0.001$, and learning rate or iteration step size to be 0.001. The implementation of ``early stop'' is to track the validation loss $l_{\text{val}}$ every $p$ steps during the training process, and if $l_{\text{val}}$ is less than the lowest loss $l_{\text{val}}^*$, then $l_{\text{val}}^*$ will be updated by $l_{\text{val}}$ and the record number $c $ is reset to be zero, otherwise, $c$ increases by 1, and when $c$ is larger than a threshold $p^*$, which means that the validation loss does not reduce for $p^*$ times, then iterations will be terminated. In our simulation, $p = 1000, p^* = 4$ are shown to produce reasonable results. The RMSprop optimizer with decay coefficient $\alpha = 0.9$ is employed to train the CNN after comparing with Adam and stochastic gradient descent methods.  Total 665 datasets (80\% are training datasets and 20\% as validating datasets) are employed to train the CNN classifier for the 39-bus power system. These datasets include the four types of faults with various fault impedance on varous load distributions. Another 560 testing datasets are generated with different load distributions. 
 
\subsubsection{Performance of the CNN under Partial Observability} 
\begin{table}[!ht]
\centering
\caption{The location accuracy rate $\eta$ ($\%$) of CNN in the 39-bus power system on the four types of faults under partial measurements } \label{39bus}
\begin{tabular}{c| c c c c }
\hline \hline 
The Ratio of Measured Buses & 15 $\%$ & 20 $\%$ & 25 $\%$ & 30 $\%$ \\ 
\hline TP & 89.5 & 	94.3 & 	94.3 & 	94.3  \\ 
\hline LG & 92.1 & 	95.7 & 	96.4 & 	96.4 \\ 
\hline DLG & 89.3 & 	92.9 & 	93.6 & 	95.0 \\ 
\hline LL &  87.1 & 	89.3 & 	95.0 & 94.3 \\ 
\hline Average & 89.5 & 93.1 & 94.8 & 95.0 \\ 
\hline 
\end{tabular} 
\end{table} 

Real-world PMU deployment is not ubiquitous. We consider scenarios where only 15\% $\sim$ 30\% of the buses are covered by PMUs. The LARs of the CNN for the 39-bus test system under partial observability are summarized in Table~\ref{39bus}. The average LARs of the four types of faults indicate that it is  easier to locate faults with more measured buses. The variations of the LARs of different faults are less than 5\% when the ratio of measured buses changes from 15\% to 30\%. The LARs are mostly higher than 90\% but less than 100\%. We further analyze these misclassified faults in the next subsection. 

\subsubsection{Analysis of Misclassified Faults} 
 \begin{table}[!ht]
 \centering 
 \caption{The misclassified faults of the CNN for the 39-bus power system under partial observability  }\label{wrong39}
 \begin{tabular}{c| c c c c }
 \hline \hline 
 Ratio of Measured Buses & 15\% & 20\% & 25\% & 30\% \\ 
 \hline within the faulted line & 89.5\% & 93.1\% & 94.8\% & 95.0\% \\ 
 \hline within \text{1-hop neighbors} & 100\% & 100\% &  100\% & 100\% \\ 
 \hline %within \text{2-hop neighbors} & 100\% & 100\% &  100\% & 100\% \\  \hline 
 \end{tabular} 
 \end{table} 
 The misclassified case happens when the line gaining the highest probability from the CNN classifier is not the faulted line. To illustrate the possible distributions of the lines gaining the highest probability, we define the following notations. 
 \begin{itemize}
 \item \textbf{``$l_{\text{prob}}$''} represent the lines with the highest probability. 
 \item \textbf{``within 1-hop neighbors''} means that the line $l_{\text{prob}}$ is directly connected with the faulted line; 
 \item \textbf{``within 2-hop neighbors''} means that at the line $l_{\text{prob}}$ and the faulted line are both connected to a common line.
 \end{itemize} 

The statistical results of all the misclassified faults are summarized in Table~\ref{wrong39}. When there are 15\% of buses are measured, in 89.5\% cases those $l_{\text{prob}}$ are exactly the faulted lines, and in 100\% cases those $l_{\text{prob}}$ are within 1-hop neighbors of the faulted line. These results demonstrate that the misclassified faults are all within the 1-hop neighbors of the faulted line under the partial observability. The main reason is that the buses have relatively small electrical distance \cite{KBLG94}.  
%\vspace{-3mm}
\subsection{Performances of 68-bus Test Cases under Partial Observability } \label{68-test}

\subsubsection{Parameters of CNN and Datasets Introduction}  
\begin{table}[!ht]
\vspace{-2mm}
\centering
\caption{The structural parameters of the CNN for the 68-bus power system }\label{CNN_para}  
\begin{tabular}{c| c c  c}
\hline \hline
Layer & Operator & Kernel  & Output \\
\hline The $1^{st}$ & Convolution & 4 @ 5 &  4 @ 64\\
 Convolutional  & Max Pooling & 2$ \times $1 &  4 @ 32 \\
\hline The $2^{nd}$ & Convolution & 8 @ 5 & 8 @ 28\\
 Convolutional  & Max Pooling & 2$ \times $1 & 8 @ 14 \\
\hline The $3^{rd}$ & Convolution & 8 @ 3 &  8 @ 12\\
 Convolutional  & Max Pooling & 2$ \times $1 & 8 @ 6 \\
\hline The $4^{th}$ & Convolution & 8 @ 3 & 8 @ 4\\
 Convolutional  & Max Pooling & 2$ \times $1 & 8 @ 2 \\
\hline Fully Connected & Vectorize & - & 16 \\
\hline Output & Regression & - & 87 \\
\hline
\end{tabular}   
%\vspace{-2mm}
\end{table}  
 The parameters of the CNN classifier for the IEEE 68-bus power systems are in Table \ref{CNN_para}. This CNN has four convolutional layers, one fully connected layer, and one output layer. The notations in Table \ref{CNN_para} follow the same principles with that in Table~\ref{CNN39}. The number of classes $m = 87$ for the CNN of the 68-bus. The weight and bias matrices in the output layer is $W_{68}^o \in R^{16 \times 87}, B_{68}^0 \in R^{87}$. The parameters of Table~\ref{CNN39} and Table~\ref{CNN_para} are selected according to the number of buses and some practical recommendations \cite{BY12}. We set the hyper-parameters to be $\lambda = 0.001, p = 1000, p^* = 4, \alpha = 0.9$, and the learning rate is 0.001. Total 1642 datasets (80\% are training datasets and 20\% as validating datasets) are employed to train the CNN classifier for the 68-bus power system. About 1210 testing datasets on various initial conditions are generated to test the performance of locating faults. 
 
 \subsubsection{Performance of the CNN under Partial Observability} 
\begin{table}[!h]
\centering
\caption{The location accuracy rate $\eta$ ($\%$) of CNN in the 68-bus power system on the four types of faults under partial measurements } \label{68bus}
 \begin{tabular}{c| c c c c }
\hline \hline 
The Ratio of Measured Buses & 15 $\%$ & 20 $\%$ & 25 $\%$ & 30 $\%$ \\ 
\hline TP & 87.3\%	 & 90.1\%	 & 91.5\%	 & 95.6\% \\ 
\hline LG & 92.1\%	 & 94.4\% & 	94.6\% & 	96.1\%  \\ 
\hline DLG & 89.5\% & 	89.2\% & 	92.0\% & 	94.9\% \\ 
\hline LL &  90.9\% & 	90.0\% & 	90.5\% & 	93.1\% \\ 
%\hline Average & 90.0\% & 90.9\% & 	92.2\% & 	94.9\% \\ 
\hline 
\end{tabular}  
\end{table} 
 \vspace{-3mm} 
\begin{table}[!ht]
\centering
\caption{The location accuracy rate $\eta$ ( $\%$) of CNN on the four types of faults with different fault impedance under different partial measurements } \label{zaver}  
 \begin{tabular}{c| c c c c }
\hline \hline 
The Ratio of Measured Buses & 15 $\%$ & 20 $\%$ & 25 $\%$ & 30 $\%$ \\ 
\hline 0.1 p.u. &  85.5\% & 	88.3\% & 	91.9\% & 	93.4\%   \\ 
\hline 0.05 p.u. & 	95.7\% & 	94.3\% & 	97.2\% & 	98.6\%  \\ 
\hline 0.01 p.u. & 92.3\% & 	90.9\% & 90.4\%	 & 93.8\% \\ 
\hline 0.001 p.u. & 93.4\% & 	94.3\% & 	94.8\% & 96.7\% \\ 
\hline 0.0001 p.u. & 87.3\% & 	88.1\% & 	87.7\% & 91.0\% \\ 
%\hline Average & 90.8\% & 		91.2\%  & 92.4\% & 		94.7\% \\
\hline 
\end{tabular}  
\end{table} 
\vspace{-3mm}
The LARs of the CNN for the 68-bus power system under partial observability are summarized in Table~\ref{68bus}. Most faults can be located with more than 90\% accuracy. The maximal variation of the LARs, due to different ratios of measured buses, is less than 10\%. The performance of locating faults of different  fault impedance are summarized in Table~\ref{zaver}. When the fault impedance is too high or too low under lower observability, the LARs are relatively low, as the proportion of these extreme cases in the training datasets are relatively small \cite{GZL17}, but with the increase of measured buses, the performances of all fault types can be improved.
% \vspace{-3mm} 
\subsubsection{Analysis of Misclassified Faults}
\begin{table}[!ht]
 \centering 
 \caption{The misclassified faults of the CNN for the 68-bus power system under partial observability  }\label{wrong}
% \vspace{-3mm}
 \begin{tabular}{c| c c c c }
 \hline \hline 
 Ratio of Measured Buses & 15\% & 20\% & 25\% & 30\% \\ 
 \hline within the faulted line & 90.0\% & 90.9\% & 	92.2\% & 	94.9\% \\ %88.7\% &  89.9\% &  92.7\% &  94.2\%  \\
% \hline IAR  (\%)& 96.3 & 94.3 & 91.6 & 87.5 \\
 \hline within \text{1-hop neighbors} & 91.3\% & 93.5\% &  96.7\% & 100\% \\ 
 \hline within \text{2-hop neighbors} & 95.2\% & 97.4\% &  100\% & 100\% \\  
 \hline 
 \end{tabular}   
 \end{table} 

 Table \ref{wrong} demonstrates the distribution of the $l_{\text{prob}}$ when the system is measured by different ratio of buses. Specifically, when 15\% of buses are measured, in 88.7\% cases $l_{\text{prob}}$ are exactly the faulted lines, in 91.3\% cases that the $l_{\text{prob}}$ are within the 1-hop neighbors of the faulted line, and 95.2\% within the 2-hop neighbors. When 30\% of buses are measured, in 94\% cases $l_{\text{prob}}$ are the faulted lines, and all the $l_{\text{prob}}$ are in the 1-hop neighbor of the faulted line. These results indicate that in more than 90\% cases $l_{\text{prob}}$ are within the 1-hop neighbor of the faulted lines, and the more buses are measured, the more likely that the $l_{\text{prob}}$ are close to the faulted lines. 
%\vspace{-3mm}
\subsection{Comparison of Different Classifiers under Partial Observability} \label{comp} 
\vspace{-2mm}
\begin{figure}[!ht]
	\centering
	\includegraphics[width=0.45\textwidth]{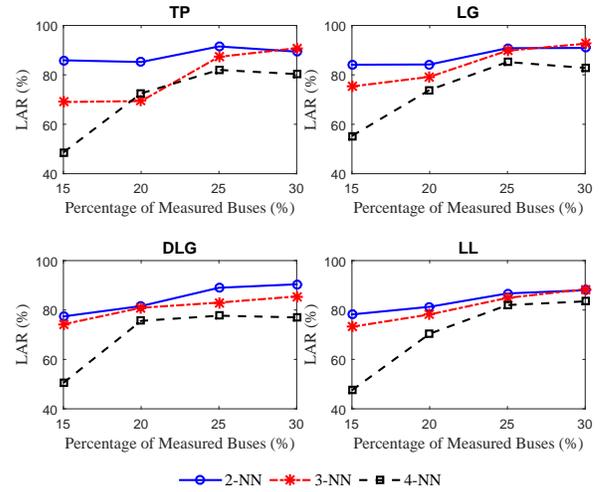} 	 	
	\caption{The LARs of NN classifier with different layer depths in terms of different percentage of measured buses} \label{NN_comp}
\end{figure}
 \begin{figure}[!ht]
	\centering
	\includegraphics[width=0.45\textwidth]{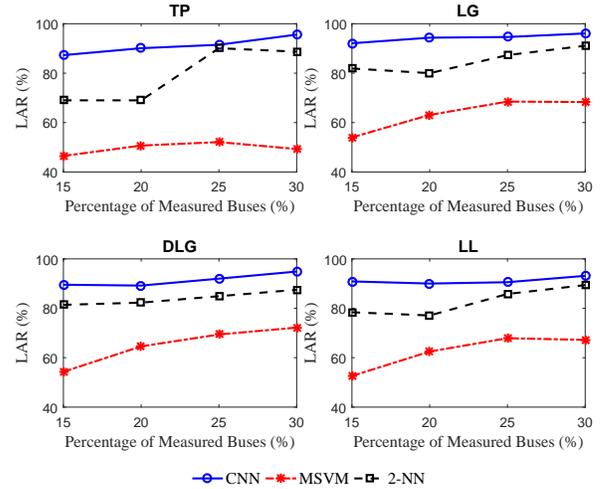} %partial_cmp LAR_cmp2.eps	 	
	\caption{ The LARs of the CNN, MSVM, NN on the four types of faults in terms of different percentage of measured buses} \label{LAR_comp}
\end{figure} 
%\vspace{-3mm}
When the system is partially observable, we compare the LARs of CNN with that of two other machine learning classifiers, including multi-class support vector machine (MSVM) \cite{PAJHH05} and ``fully-connected'' neural network (NN). The MSVM classifier is based on the coupling pairwise method with the radial basis function kernel to find the global solution. NN of two $ \sim $ four layers are tested, and the two-layer NN is selected as it achieves the optimal performance as shown in Fig.~\ref{NN_comp}, which demonstrates that the two-layer NN has better performance than other schemes. The weight and bias matrices for the first layer of NN are $W^1_{NN} \in R^{68 \times 32}, b^1_{NN} \in R^{32}$ and for the second layer are $W^2_{NN} \in R^{32 \times 16}, b^2_{NN} \in R^{16}$, and the activation function is ReLU function $f(x) = \max(x,0)$.  We set the hyper-parameters to be $\lambda = 0.001, p = 1000, p^* = 4, \alpha = 0.9$, and learning rate is 0.001. Notice that here the NN has a smaller number of hidden units than the dimension of output. The performances of NNs with an equal or greater number of hidden units are also tested and compared, and the average performances of all these NNs are similar and less than that of CNN. %The RMSprop optimizer with decay coefficient $\alpha = 0.9$ is employed to train both NN. gradient descent methods. 

Under such partial observability, the LARs of the MSVM, two-layer NN and CNN are compared for the four types of faults in Fig.~\ref{LAR_comp}. The observed buses for each classifier are selected according to the principles of algorithm \ref{alg} using their corresponding loss functions to demonstrate optimal performance. 
 
The results in Fig.~\ref{LAR_comp} demonstrate that when only 15\% $\sim$ 30\% buses are observed, fault localization by CNN is much better for the four types of faults than that shown by the other two classifiers. Observe that when 30\% of buses are measured, CNN can reach an impressive fault localization accuracy of more than 95\% for faults of the four types. %{\color{red} In addition, the proposed method is also implemented in the IEEE 39-bus power system, and the performance is similar when there are 15\% $\sim$ 30\% buses measured, and the } 
\subsection{The ARC of CNN under $\leq$ 15\% of nodal observability}\label{arc}
It is worth investigating the performance of the CNN classifier when less than 15\% of all buses are measured. In this case, one would guess that LARs of CNN cannot be better than 90\%. However we observe that even if the CNN does not predict the fault location accurately, it is still able to associate a relatively large probability of failure (though not the largest) to the correct faulted line. To analyze this, we sort the lines according to the output probability $\bar{y}^j$ of CNN in descending order and then record the rank $r_j$ of the correct line of the $j$th fault. We define a new performance metric ``average rank of the correct line'' (\textbf{ARC}) for the $N$ testing faults as $\bar{r} = \frac{1}{N} \Sigma_{j=1}^N r_j$. 

The ARC indicates how many high-probability lines need to be considered on average to show the correct faulted line. Note that a lower ARC reflects better average performance with the ARC of exact localization being $1$.

\begin{table}[!ht]
\centering
\caption{The ARC of CNN for different type of faults when the ratio of measured buses is less than 15\% }\label{arc_cmp}
\begin{tabular}{c| c c c c }
\hline \hline
 Measured Ratio & TP & LG & DLG & LL \\
\hline 7\% & 1.32 & 1.48 & 1.92 & 1.56 \\
\hline 10\% & 1.38 & 1.28 & 1.66 & 1.54 \\
\hline 15\% & 1.38 & 1.23 & 1.57 & 1.54 \\
\hline
\end{tabular}
\end{table}
The ARC of the four types of faults is shown in Table \ref{arc_cmp} when no more than 15\% of buses are measured. It is significant that the ARC for all types of faults is less than $3$ when only 7\% to 15\% of buses are measured. This observation suggests that despite the low PMU coverage, the operator needs to check only a few lines to identify the fault. Crucially, as discussed next, under low PMU coverage, CNN is also able to localize the fault to a small graphical neighborhood of its true location.
\vspace{-3mm}
\subsection{Neighborhood property of high probability lines}
 \label{neighbor}
\begin{figure}[!ht]
	\centering
	\includegraphics[width=0.35 \textwidth]{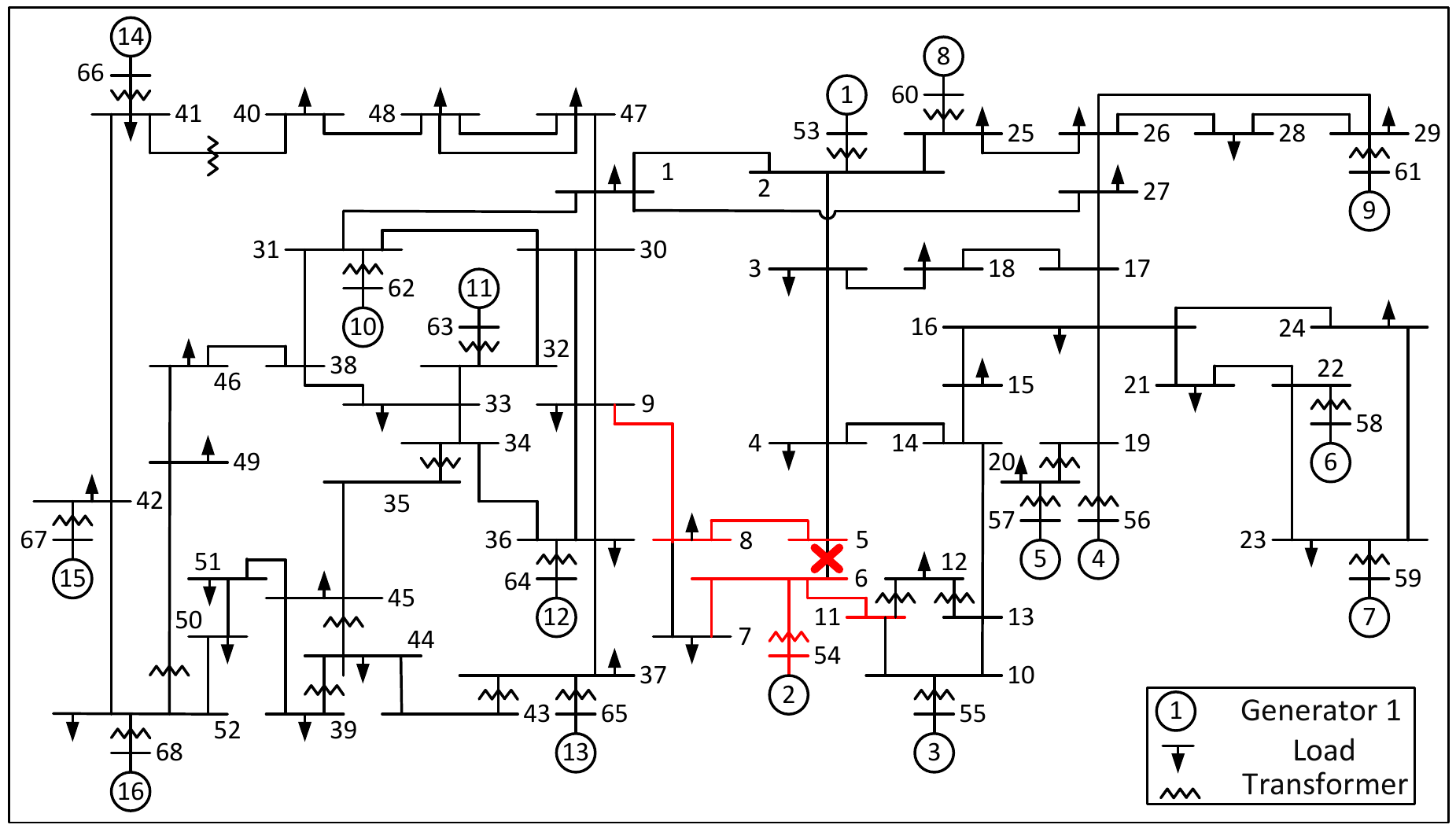} 	 	
	\caption{The lines of top-5 high probability (above) from CNN are marked in red IEEE 68-bus power system } \label{top5}
	\vspace{-3mm}
\end{figure}
The lines with high output probability $\bar{y}_i^j$ demonstrate neighborhood property in Fig.~\ref{top5}, where the line between bus 5 and 6 has a three-phase short circuit fault. All lines are sorted according to $\bar{y}_i^j$ from high to low, then those with the top-5 probabilities, marked as red, are in the neighborhood of the faulted line. Furthermore, we have verified that this neighborhood property is not a special case for this fault but extends to the majority of the tested faults. Moreover, this neighborhood property is determined by the feature vector in \eqref{neighb} and as such also applies to other tested classifiers, e.g. NN. Since, $\psi_k (k\neq i, j )$, defined in Section \ref{interpret} as the total line currents in the neighborhood of bus $k$, lines in the neighborhood of the fault are identified with high probability.

Low ARC and neighborhood localization properties appear very useful to guide initial dispatch of a recovery/maintenance crew. Moreover, it should also be advantageous to use these features to determine the order of triggering relays or circuit breakers automatically for protection in the post-fault grid. We plan to study these directions in the future. 
\subsection{Comparison with other PMU placement algorithms}\label{cmp_pmu}
\begin{figure}[!h]
\vspace{-3mm}
	\centering
	\includegraphics[width=0.48\textwidth]{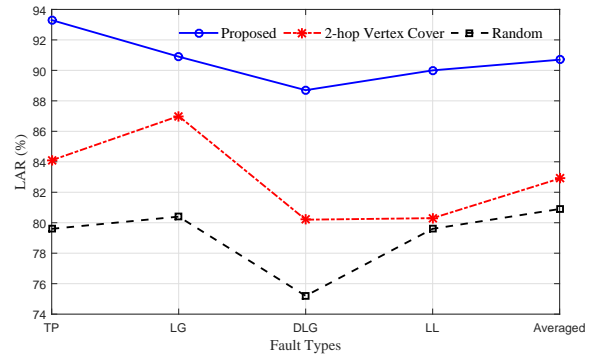} 	 	
	\caption{The LARs of CNN when 12 measured buses are selected by three algorithms} \label{alg_cmp}
	\vspace{-3mm}
\end{figure}

In this Subsection, we discuss the performance of the algorithm \ref{alg} for PMU placement. The proposed algorithm is compared with the ``2-hop Vertex Cover (VC)'' and the Random placement algorithms. The ``2-hop VC'' is a topology-based algorithm for PMU placement \cite{deka2011pmu}. It places PMUs on a set of buses such that each edge in the graph is at most two hops away from a PMU. The baseline of the Random algorithm selects arbitrarily $s$ buses. The LARs for faults of the four tested types are compared in Fig.~\ref{alg_cmp} where the measured buses are suggested by the three placement algorithms.

As there are at least 12 buses that can satisfy the objective of ``2-hop vertex cover'' for this 68-bus power system, these three algorithms are compared when $s=12$. The 12 buses selected by the Random algorithm include $\S_{R}=[31,3,65,46,43,28,15,44,23,58,9,57]$, one solution of the 2-hop VC algorithm is obtained by solving a linear programming approximating the 2-hop VC formulation, and the selected buses are $\S_{VC}=[3,6,13,19,23,26,30,31,36,40,44,52]$, and the 12 buses selected by the method proposed in the manuscript are $\S_P=[1,9,16,30,36,23,42,61,51,57,6,37]$. %{\color{red} These buses are selected as the training loss $l =$ is the the lowest while the training loss based on $\S_R$ and $\S_{VC}$ are ?, ? respectively. } 
Compared with the Random algorithm, the improvements of the proposed algorithm for the different types of faults varies, but it shows about 10\% improvement in average over the other methods. The 2-hop VC method also has higher LARs than that of the Random algorithm, but it is still lagging behind the proposed algorithm showing the average improvement of 8\%.

 The running time of the proposed PMU placement algorithm is 5.4 hours based on the personal computer of Intel i7, 3.6 GHz CPU and 32 GB RAM, and this process can be accomplished offline. The running time of testing one dataset is 4 ms, thus the proposed CNN classifier can efficiently determine the location of the faults in real-time.  
 \vspace{-3mm}
 \subsection{Robustness to Uncertainty of Loads}\label{load}
 \vspace{-3mm}
 \begin{table}[!ht]
 \centering 
 \caption{The LARs of all types of faults tested on the datasets with different uncertainty index $\zeta $ }\label{zeta}
% \vspace{-3mm}
 \begin{tabular}{l| c c c c c}
 \hline \hline 
 Fault Type & TP & LG & DLG & LL & Average\\ 
 \hline $ \zeta $ = 0.13 & 95.4 & 	99.4 & 	98.9 & 	98.9 & 98.2 
 \\ 
 \hline $ \zeta $ = 0.38 & 95.4 & 	96.1 & 	94.9 & 	93.1 & 94.9\\ 
 %\hline $ \zeta $ = 5.9  & 92.9 & 	96.6 & 	93.8 & 93.4 & 	 94.2 \\ 
 \hline $ \zeta $ = 0.51  & 98.0 & 	91.6 & 	90.4 & 	84.1 & 	 91.0 \\
 %\hline $\zeta=17.6 $ &  \\ 
 \hline 
 \end{tabular} 
 \vspace{-3mm}
 \end{table}
  
The main uncertainty in the power system comes from the randomness of load fluctuations and noise, which cause the difference between training and testing datasets. To measure the uncertainty caused by the load fluctuations, we define uncertainty index $\zeta$ 
\begin{align} 
\zeta & = \frac{1}{nN'} \Sigma_{i=1}^{N'} \frac{\lVert U^{0,i} - \bar{U}^0 \Vert_2 }{\lVert \bar{U}^0 \Vert_2 } 
\end{align}
where $N'$ is the number of datasets, $n$ is the number of buses, and $U^{0,i} \in \C^{n \times 1}$ is the pre-fault bus voltages of the $i$th dataset, and $\bar{U}^0 $ is the mean value of the pre-fault bus voltages of  all the training datasets. 

$\zeta$ denotes the averaged differences of individual bus voltages between the testing and the training datasets. A larger $\zeta$ indicates that the datasets have large variances or higher complexity. To illustrate the robustness of the classifier to such uncertainty, the location accuracy rates are compared with the different degree of load variations in Table \ref{zeta}. Here the results of 3630 datasets with different load fluctuations are illustrated based on 30\% measured buses. The uncertainty index $\zeta$ of the testing datasets changes from 0.13 to 0.51.  
 
 The results in Table\ref{zeta} demonstrate that the classifier can tolerate some load variations, and the LARs are more than 90\% if the $\zeta$ is less than 0.4. When $\zeta$ is small, indicating the load variations are mild. When $\zeta$ is 0.13, the average LAR is close to 100\%, and when $\zeta$ increases to 0.51, the average LAR is less than 95\%. Notice that for training datasets $\zeta$ = 0.20. Thus when $\zeta$ is far more than 0.2, the classifier need to be updated by adding more new training datasets. We also confirmed that the average LAR has about 6\% improvement after adding another more than 1000  training datasets with larger $\zeta$ to retrain the CNN classifier.
 \vspace{-3mm} 
 \subsection{Robustness to noise}\label{noise}
% \vspace{-5mm}
 \begin{figure}[!h]
% 	\vspace*{-3mm}
 	\centering
 	\includegraphics[width=0.24\textwidth]{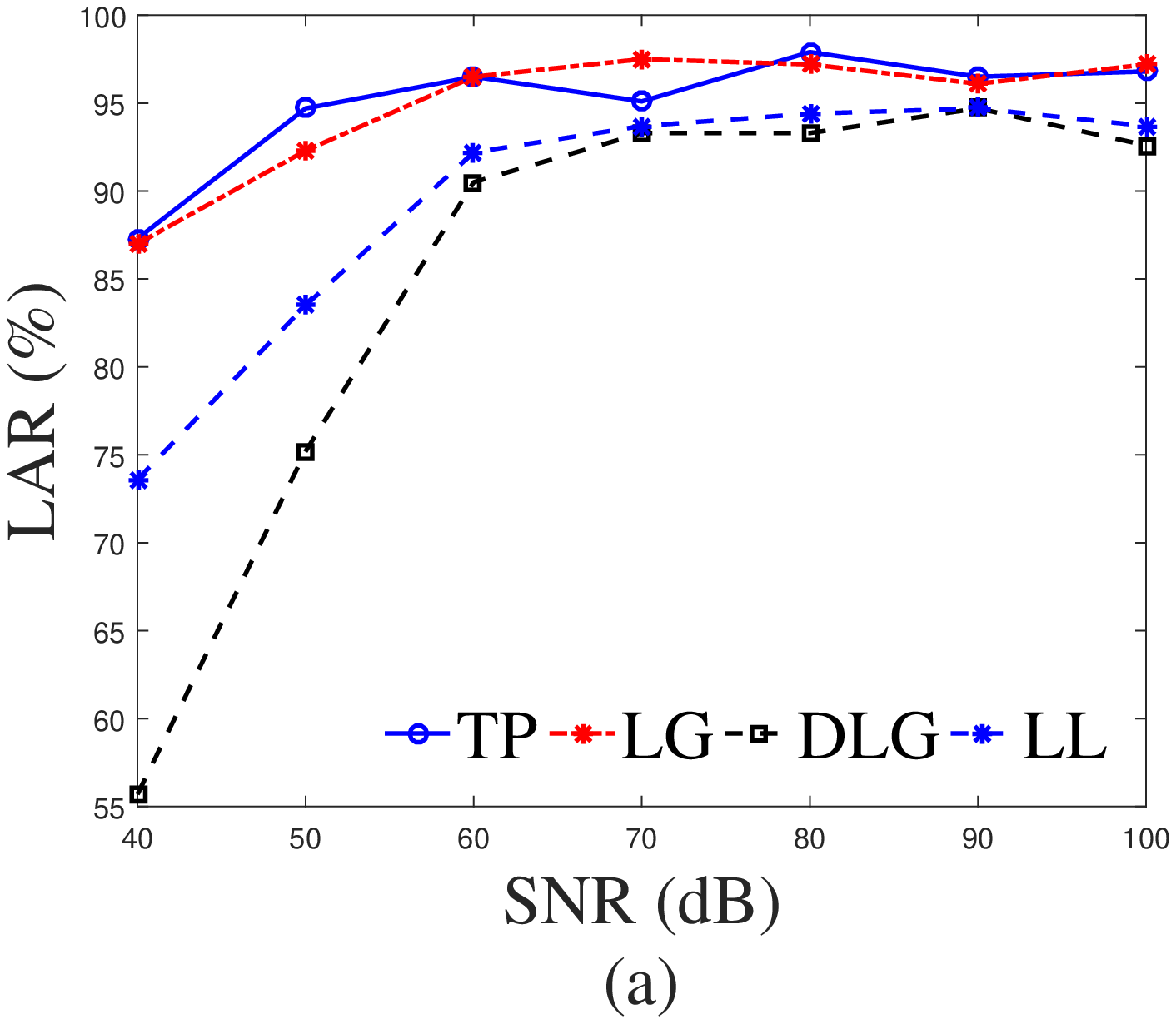}%Figure_noise_a.eps 
 	\includegraphics[width=0.24\textwidth]{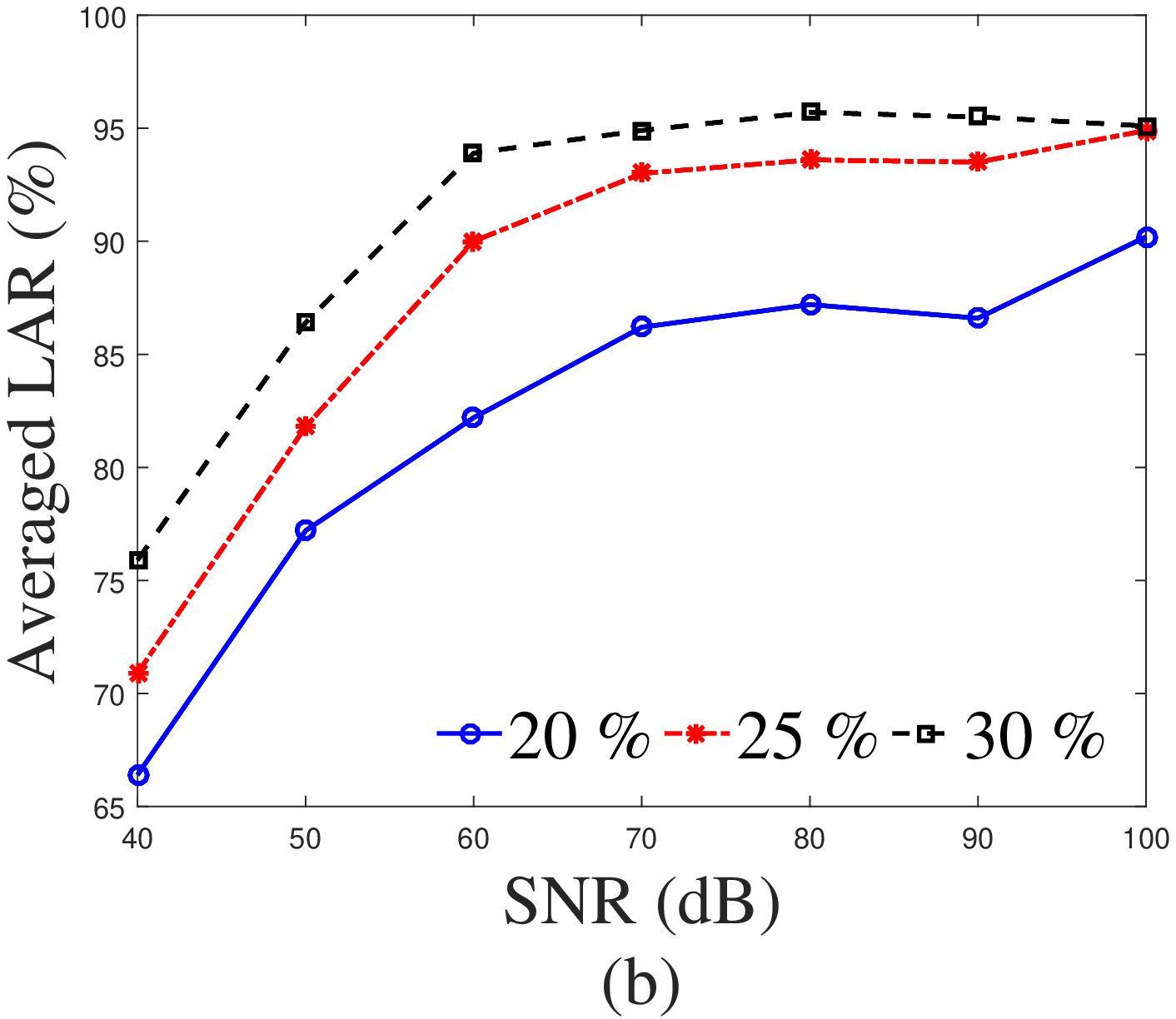}%Figure_noise 
 	\caption{(a) The LARs for the different types of faults when 30\% of buses measured with different SNR; (b)The Averaged LARs of all types of faults when 20\% $\sim$ 30\% of buses measured with different SNR} \label{noise1}
 	\vspace{-2mm}
 \end{figure}
The IEEE Standard C37.118 only defines the measurement accuracy but does not specify the signal-noise-ratio(SNR) of PMU measurements \cite{MK15}, and the SNR of PMUs in different regions can vary. We select the experimental range of SNR from 40 dB to 100 dB \cite{BBBRCH16,XCLR14, LWC18}. Gaussian noise of the same SNR is added both to the training and testing datasets. The structure of the CNN is the same as before but the hyper-parameter decay coefficient, $\alpha$, is changed from $0.9$ to $0.7$ in the noisy regime. Other parameters are the same.
 
 Fig.~\ref{noise1} (a) demonstrates the LARs with different SNR when 30\% of buses are observed, and the Fig.~\ref{noise1} (b) shows the average LARs of all types of faults when 20\% $\sim$ 30\% of buses are observed. Results in Fig.~\ref{noise1} (a) indicate that the sensitivity of different types of faults to noise is different, and the three-phase short circuit faults are relatively more robust to the noise. When SNR is higher than 60 dB, LARs for all the types of faults can achieve 90\% or higher. Fig.~\ref{noise1} (b) reveals that, as expected, when more buses are measured the robustness to noise can be strengthened. Furthermore, when SNR is higher than 60 dB, the influence of the noise is contained, and the performance does not improve or degrade noticeably. 
% \vspace{-3mm}
\subsection{The Impacts of Voltage Quality} \label{quality}
During the fault period, the viability of voltage measurements are considered for practical implementation. The impacts of inaccuracy of synchronization and the complex transients states due to different fault clearing time are studied. 
\vspace{-3mm}
\subsubsection{The Impacts of inaccuracy of synchronization}    
\begin{table}[!ht]
\vspace{2mm} 
\centering 
\caption{The differences $\nu_d$ of the delayed LAR and the normal LAR for the different mean $\mu_d$ of the distribution of the delay time  }\label{delay}  
\begin{tabular}{l| c c c  }
\hline \hline 
$\mu_d$ (ms) & 20 & 30 & 40 \\ % & 50 
\hline $\nu_d$ of TP ($\%$) & 0.1 & 0.6 & 0.6  \\ % & 0.9 
\hline $\nu_d$ of LG ($\%$) & 0 & 2.8 & 3.2  \\ %& 4.8 
\hline $\nu_d$ of DLG ($\%$) & 0.3 & 3.4 & 4.8 \\ %& 9.6
\hline $\nu_d$ of LL ($\%$) & 0.5 & 2.5 & 2.5  \\ %& 7.3 
\hline 
\end{tabular} 
\vspace{-2mm}
\end{table}  
According to the standards of PMUs manufacture in IEEE C37.118.1-2011, it is required the measured voltage, currents, and frequency to satisfy the 1\% maximum Total Vector Error (TVE), and the PMU measurement delay should be within the dynamic requirements \cite{MK15}. The delays can be characterized by the Gaussian distribution, and the suggested mean is 20 milliseconds (ms) with standard deviation of 6 ms depending on variant communication systems \cite{ZK13}. Thus we randomly select some of the measured buses with the delayed measurements following the Gaussian distribution $\mathcal{N}(\mu_d, \sigma_d)$, where $\sigma_d = $ 6 ms and $\mu_d \in \{20, 30, 40\}$.

Specifically, we generate datasets with the simulation step of 0.001 second and down-sample the datasets to have the data rate around 60 samples per second. There are 50\% buses randomly selected with delayed measurements while others are with the exact measurements, then we employ  the accurate and delayed measurements to compute the feature vector $\bar{\psi}$. The resulting LAR is defined as delayed LAR, and the difference $\nu_d$ = (LAR - delayed LAR) is calculated to measure the influence of the delayed measurements. The $\nu_d$ of the four types of faults in terms of different $\mu_d$ are summarized in Table\ref{delay}. Notice that these results are produced when there are 30\% buses are measured. 
 
The Table\ref{delay} shows the robustness of the proposed method to different levels of delay. When the delay is around 20 ms, the location performance is almost not impacted as $\nu_d$ is less than 1\%, and when data have a longer time delay,  $\nu_d$ is within 5\% when $\mu_d = 30 \sim 40$ ms.  

\subsubsection{The Impacts of fault clearing time}
\begin{table}[!ht]
\centering 
\caption{The LAR difference $\nu_f$ due to variant fault clearing time }\label{clear}
 \begin{tabular}{l| c c c c }
\hline \hline 
$T_f$ (s) & 0.05 & 0.1 & 0.15 \\ 
\hline $\nu_f$ of TP ($\%$) & 0.5 & 1.4 & 0.5  \\ 
\hline $\nu_f$ of LG ($\%$) & 0 & 0 & 0.3 \\ 
\hline $\nu_f$ of DLG ($\%$) & 1.1 & 0.6 & 0.6 \\
\hline $\nu_f$ of LL ($\%$) & 0 & 0 & 0.3 \\ 
\hline 
\end{tabular} 
\end{table}
The fault transients can be variant due to different fault clearing time, so the performance of the proposed method is tested when the fault clearing time $T_f$ changes from 0.05 s to 0.2 seconds. Here 30\% buses are measured. For the training datasets $T_f = 0.2$ seconds. The LAR with different fault clearing times is denoted as $\eta_f$. Then the differences $\nu_f = \lvert \eta - \eta_f\rvert$ for variant fault clearing time are illustrated in Table \ref{clear}. The $\nu_f$ of different types of faults is   less than 2\%, demonstrating the robustness to different fault clearing time. 
 
In addition, phase-angle jump\footnote{Phase-angle jump means a sudden change in phase angle during a short circuit. This phenomenon is due to the different ratio of reactance $X$ to impedance $R$ between the source and the feeder \cite{BBH00}.} is also possible to cause the voltage sag depending on the degree of the jumped angle. In the distribution system, phase angle jump can be tens of degree but is much smaller in the transmission system \cite{RHR99}. The averaged phase angle jump in our testing system is around ($\ang{10} \sim \ang{20}$) for the different fault impedance. Thus the proposed method is robust to such phase angle jump, but the application to the distribution system needs to consider the more significant phase angle jump up to $\ang{60}$, which will be our future research. 
% \vspace{-5mm}
\section{Conclusions}\label{con} 
This manuscript proposes to locate the faulted line by a CNN classifier through the use of physically interpretable feature vectors. The clear physical interpretations of these feature vectors are explained based on the substitution theory. The benefits of employing these features have three aspects: 1) a high location accuracy rate is reached even when the system has low observability; 2) the true faulted line, even if not the highest, has a high output probability (among top-k, if not the highest) from the classifier in low observability; 3) the misclassified lines output by the classifier mostly are geographically near the faulted line. The location performance is further increased by using the design of the CNN classifier into a joint PMU placement algorithm, that is demonstrably superior to other random and topology-based methods. The significance of the proposed faulted line localization method over other methods is revealed in the robustness to different conditions, including variant random load fluctuations, noise, inaccurate voltage measurements in the IEEE 39-bus and 68-bus power systems. 

In the future, we will extend this work to the distribution system and identify the exact location of the fault along the line. Furthermore, testing the methodology on real-data (as opposed to synthetically generated data) is another direction for our future work. 
 
%\section*{Acknowledgment}
%The authors acknowledge the support from the Department of Energy through the Grid Modernization Lab Consortium and the Center for Non-Linear Studies (CNLS) at Los Alamos National Laboratory.
\bibliographystyle{IEEEtran}
\bibliography{./IEEEabrv,./ref,./MengWangPub}
% 	 \vspace{-6.9in} 
 \begin{IEEEbiography} 
	 [{\includegraphics[width=1.05in,height=0.9 in,clip,keepaspectratio]{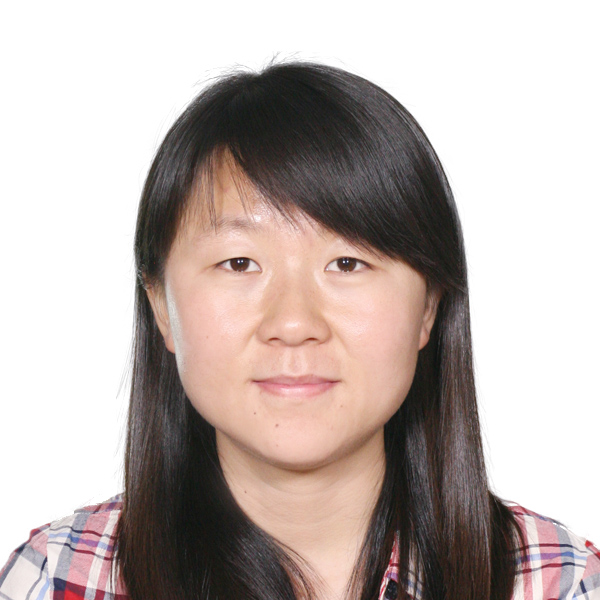}}]	 	 
	 {Wenting Li} (S'16) received the B.E. degree from Harbin Institute of Technology, Heilongjiang, China, in 2013. 
	 
	 She is pursuing the Ph.D. degree in electrical engineering at Rensselaer Polytechnic Institute, Troy, NY. Her research interests include  high-dimensional data analytics, feature extraction,  application of machine/deep learning methods to identify and locate events in power grids. 
	 	 \end{IEEEbiography}
	 \vspace{-7in} 	 
 \begin{IEEEbiography}
  [{\includegraphics[width=1.05in,height=1.25in,clip,keepaspectratio]{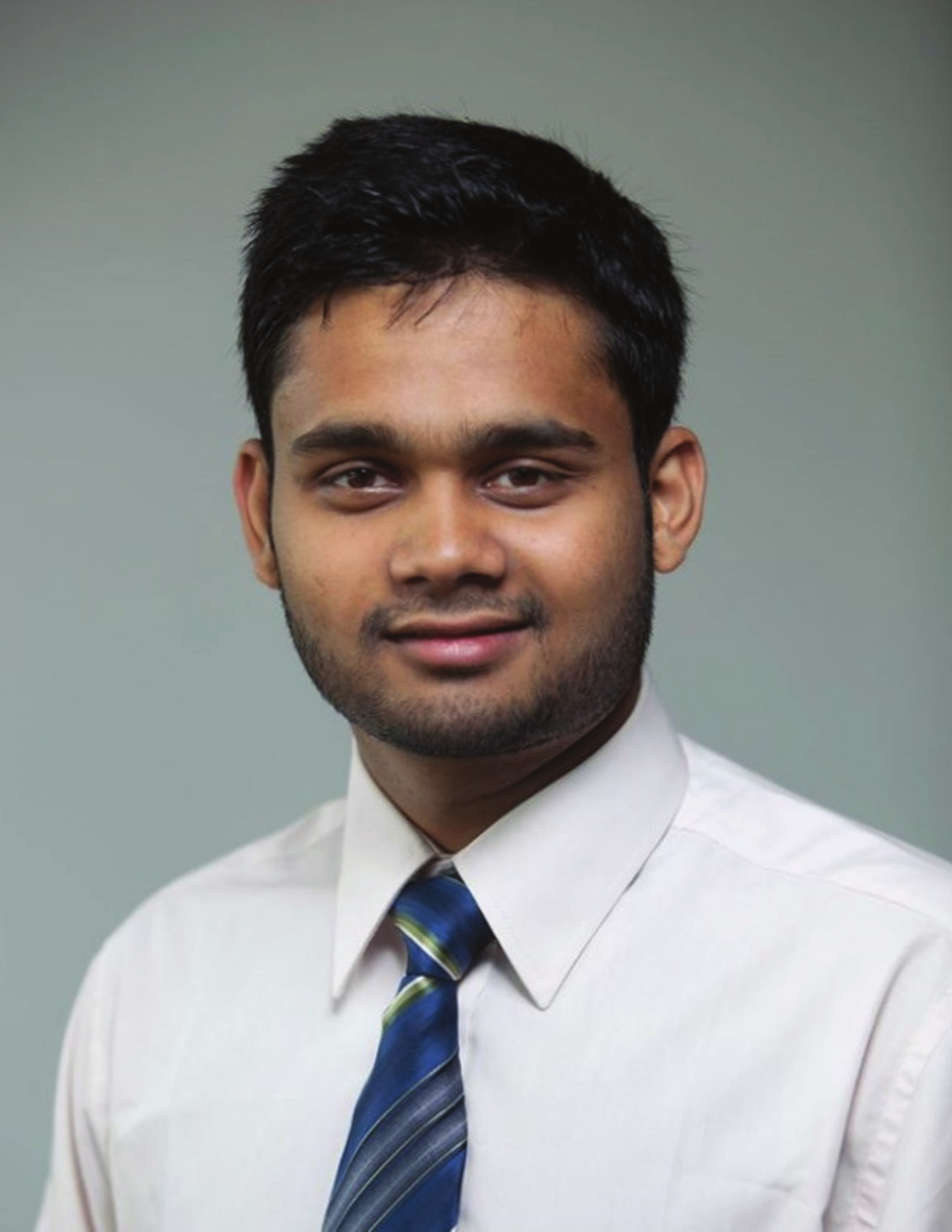}}] 
{Deepjyoti Deka} (M'11) is a staff scientist in the Applied Mathematics and Plasma Physics group of the Theoretical Division at Los Alamos National Laboratory, where he was previously a postdoctoral research associate at the Center for Nonlinear Studies. His research interests include data-analysis of power grid structure, operations and security, and optimization in social and physical networks. At LANL, Dr. Deka serves as a co-principal investigator for DOE projects on machine learning in distribution systems and in cyber-physical security. Before joining the laboratory he received the M.S. and Ph.D. degrees in electrical engineering from the University of Texas, Austin, TX, USA, in 2011 and 2015, respectively. He completed his undergraduate degree in electrical engineering from IIT Guwahati, India  in 2009 with an institute silver medal as the best outgoing student of the department.
 \end{IEEEbiography}
 	 \vspace{-7in} 
	 \begin{IEEEbiography}
	  [{\includegraphics[width=1in,height=1.25in,clip,keepaspectratio]{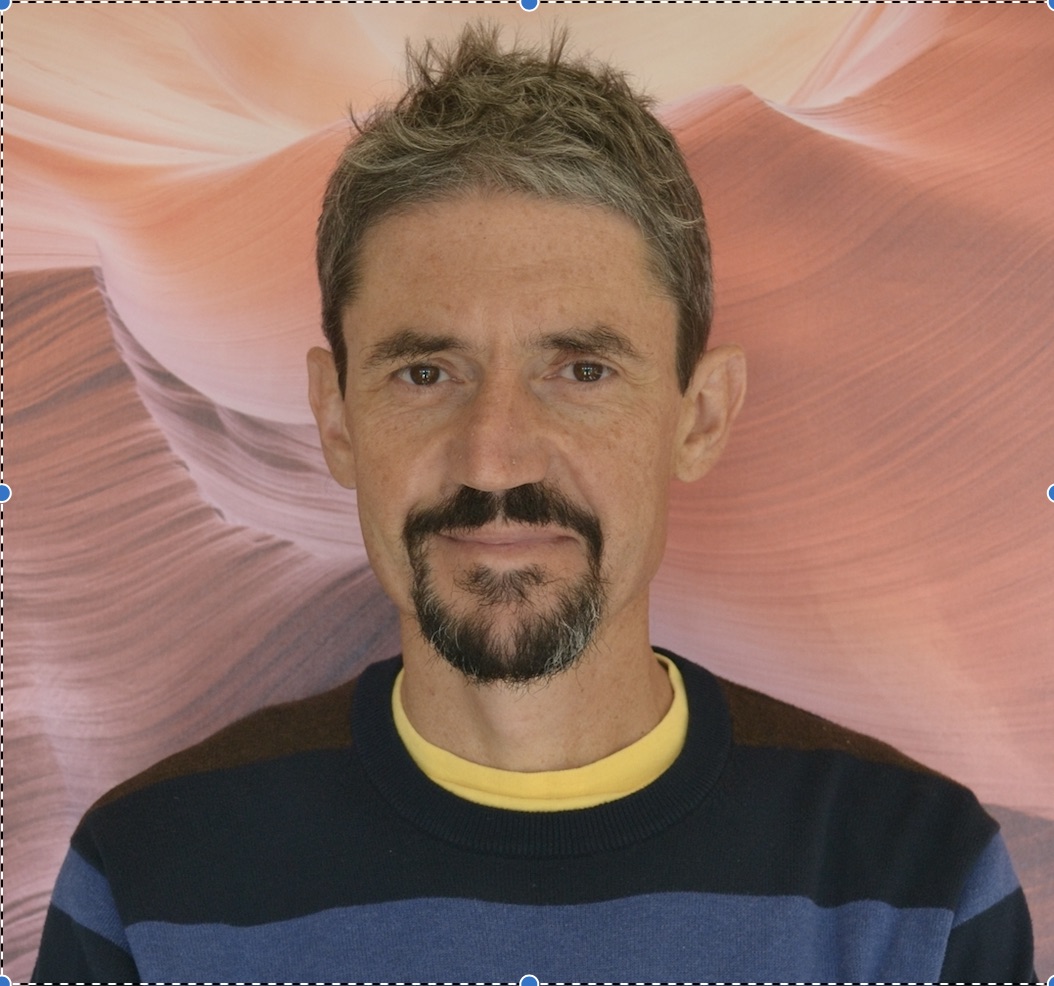}}]
	 {Michael Chertkov} Prof. Chertkov's areas of interest include mathematics and statistics applied to physical, engineering and data sciences. Dr. Chertkov received his Ph.D. in physics from the Weizmann Institute of Science in 1996, and his M.Sc. in physics from Novosibirsk State University in 1990. After his Ph.D., Dr. Chertkov spent three years at Princeton University as a R.H. Dicke Fellow in the Department of Physics. He joined Los Alamos National Lab in 1999, initially as a J.R. Oppenheimer Fellow in the Theoretical Division, and continued as a Technical Staff Member leading projects in physics of algorithms, energy grid systems, physics and engineering informed data science and machine learning for turbulence. In 2019 Prof. Chertkov moved to Tucson to lead Interdisciplinary Graduate Program in Applied Mathematics at the University of Arizona, continuing to work for LANL part time.  Dr. Chertkov has published more than 200 papers. He is a fellow of the American Physical Society (APS) and a senior member of IEEE. 
	 \end{IEEEbiography} 
	 \vspace{-7in}  
	 \begin{IEEEbiography}
	  [{\includegraphics[width=1in,height=1.25in,clip,keepaspectratio]{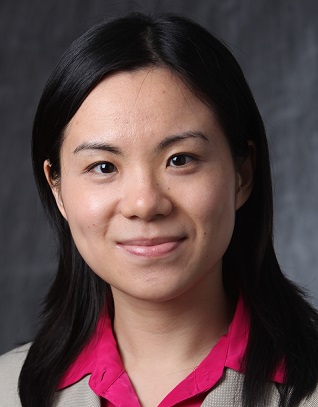}}]
	 {Meng Wang} (M'12) received B.S. and M.S. degrees from Tsinghua University, China, in 2005 and 2007, respectively. She received the Ph.D. degree from Cornell University, Ithaca, NY, USA, in  2012. 
	 
	 She is an Assistant Professor in   the department of  Electrical, Computer, and Systems Engineering at Rensselaer Polytechnic Institute, Troy, NY, USA. Her research interests include high-dimensional data analytics, machine learning,  power systems monitoring, and synchrophasor  technologies. 
	 \end{IEEEbiography}   
	 \vspace{-4 in}

\end{document}